\documentclass[aps, twocolumn,amsmath,amssymb,floatfix,superscriptaddress]{revtex4-1}
\usepackage[table]{xcolor}
\usepackage{graphicx}
\usepackage{amsmath,fullpage,amssymb}
\usepackage{graphics,epsfig}
\usepackage{makeidx}

\def\rmd{{\mathrm{d}}}

\def\rme{{\mathrm{e}}}
\def\lB{\ell_{\mathrm{B}}}

\def\Eq{Eq.}
\def\Eqs{Eqs.}
\def\Figure{Figure}
\def\Fig{Fig.}
\def\Figs{Figs.}

\def\ie{{\em i.e.}}
\def\eg{{\em e.g.}}

\newcommand{\chgA}[1]{\textcolor{black} {#1}}

\newcommand{\Av}[1]{{\bf #1}}

\newcommand{\kB}{k_\mathrm{B}}
 
\newcommand{\trm}[1]{{\textrm{#1}}}

\begin{document}

\title{Ionic structure around polarizable \chgA{metal} nanoparticles in aqueous electrolytes}
\author{Bendix Petersen}
\affiliation{Research Group for Simulations of Energy Materials,
Helmholtz-Zentrum Berlin f\"ur Materialien und Energie,
  Hahn-Meitner-Platz 1, 14109 Berlin, Germany}
\affiliation{Institut f\"ur Physik, Humboldt-Universit\"{a}t zu Berlin, Newtonstr.\ 15, D-12489 Berlin, Germany}
\author{Rafael Roa}
\affiliation{Research Group for Simulations of Energy Materials,
Helmholtz-Zentrum Berlin f\"ur Materialien und Energie,
  Hahn-Meitner-Platz 1, 14109 Berlin, Germany}
\affiliation{Departamento de F\'{i}sica Aplicada I, Facultad de Ciencias, Universidad de M\'{a}laga, Campus de Teatinos s/n, E-29071 M\'{a}laga, Spain}
\author{Joachim Dzubiella}
\affiliation{Research Group for Simulations of Energy Materials,
Helmholtz-Zentrum Berlin f\"ur Materialien und Energie,
  Hahn-Meitner-Platz 1, 14109 Berlin, Germany}
\affiliation{Institut f\"ur Physik, Humboldt-Universit\"{a}t zu Berlin, Newtonstr.\ 15, D-12489 Berlin, Germany}
\author{Matej Kandu\v c}
\email{matej.kanduc@helmholtz-berlin.de}
\affiliation{Research Group for Simulations of Energy Materials,
Helmholtz-Zentrum Berlin f\"ur Materialien und Energie,
  Hahn-Meitner-Platz 1, 14109 Berlin, Germany}


\pagenumbering{arabic}
\noindent

\parindent=0cm
\setlength\arraycolsep{2pt}

\begin{abstract}
Metal nanoparticles are receiving increased scientific attention owing to their unique physical and chemical properties that make them suitable for a wide range of applications 
in diverse fields, such as electrochemistry, biochemistry, and nanomedicine.
Their high metallic polarizability is a crucial determinant that defines their electrostatic character in various electrolyte solutions.
Here, we introduce a continuum-based model of a metal nanoparticle with explicit polarizability in the presence of different kinds of electrolytes.
We employ several, variously sophisticated, theoretical approaches, corroborated by Monte Carlo simulations in order to elucidate the basic electrostatics principles of the model. 
We investigate how different kinds of asymmetries between the ions result in non-trivial phenomena, such as charge separation and a build-up of a so-called zero surface-charge double layer.
\end{abstract}
\maketitle
\setlength\arraycolsep{2pt}

\section{Introduction}

The field of electrostatic interactions in classical soft-matter and biological systems has a long and rich history, recognized by many intellectual challenges and ideas.~\cite{Linse2, holm2012electrostatic, dean2014electrostatics}
Maybe one of the most remarkable concepts is the approximation of implicit solvent, a continuum-level approach where a system composed of charged species and solvent molecules is simply treated as a gas of the charged species only, now with their interactions attenuated by the relative dielectric permittivity $\varepsilon$ of the solvent.~\cite{bottchertheory}
Remarkably, this so-called {\it primitive model} or {\it dielectric approximation} works very well for simple ions in aqueous solutions down to only several layers of water molecules between the ions.~\cite{gavryushov2006effective, hess2006modeling}
This is also one of the reasons for the efficiency of the Poisson--Boltzmann equation to describe monovalent ions in the water environment.
Nonetheless, an implementation of this approximation can become technically involved
if dielectric discontinuities are present in the system. This is unfortunately also one of the reasons why dielectric discontinuities are commonly neglected in (too) many studies. Neglecting them is not always justified, because in the presence of charges, a dielectric discontinuity leads to a polarization surface charge density at the boundary, which \chgA{further} influences the local electrostatic potential and interactions with surrounding charges. 

In the case of a planar dielectric discontinuity, the electrostatic potential can simply be expressed as the electrostatic potential arising from a fictive ``image charge'' residing on the other side of the discontinuity. Therefore, in this context, the dielectric discontinuity effects are sometimes referred to as {\it image charges}. 
The dielectric effects in double-layer problems of planar geometry have been elaborated by Torrie, Valleau, and Patey,~\cite{II} and by Bratko, Jonsson, and Wennerstr\"{o}m~\cite{bratko1986electrical} using computer simulations, and by Kjellander and Mar\v{c}elja~\cite{kjellander1984correlation} and Outhwaite, Bhuiyan, and Levine~\cite{outhwaite1980theory} utilizing various theoretical frameworks.
The image-charge concepts have been adapted to spherical symmetry by Linse.~\cite{Linse1, Linse2, greenfunction2} He showed that approximating the exact mathematical expressions for the spherical geometry leads to a simplified picture in which the polarization is described by image charges as in planar cases.
The image charges in the spherical geometry are of paramount importance, since a vast majority of the soft-matter electrostatics research in the recent decades has \chgA{focused} on colloidal and biological systems, where various macromolecular structures (\eg, colloids, proteins, polysaccharides, micelles) in water can be modeled as spherical entities with a lower dielectric interior $\varepsilon'$ (due to their predominantly hydrocarbon architectures) than the surrounding water environment ($\varepsilon'\ll\varepsilon$).~\cite{Linse2,messina, zwanikken2007charged, rescic2008potential, overcharging, greenfunction, bakhshandeh2011weak, pryamitsyn2015pair, effectsJCP,dos2016simulations, inhomogeneous, delacruz, tergolina2017effect}

The other side of the spectrum, containing spherical bodies of a much higher dielectric interior than water (\ie, $\varepsilon'\gg\varepsilon$), such as for example small metal particles in aqueous environments, has been much less explored.
However, the interest in this field has boosted with recent advances in metal nanoparticle chemistry and physics, which have emerged as a broad new discipline in a subdomain of colloids and surfaces.~\cite{AuNP, sardar2009gold}
One of the most prominent discoveries was that gold nanoparticles (of a size 1--10~nm) are active catalysts for oxidation reactions.~\cite{haruta1997size} This has triggered a tremendous research activity in nanocatalysis, which presently remains one of the fastest growing areas of nanoscience.~\cite{burda2005chemistry, ferrando2008nanoalloys, catalysis}
Furthermore, applications involving metal nanoparticles can for instance be found in electrochemistry for nanoelectrodes,~\cite{welch2006use} photovoltaic cells~\cite{beek2004efficient} electro-osmosis,~\cite{bazant2004induced} or in biochemistry and nanomedicine for drug delivery, therapeutics, diagnostics, and bioimaging.~\cite{chung2007effect, asharani2008cytotoxicity, lin2009synthesis, ackerson2010synthesis, bowman2008inhibition}
 At the same time, experimental findings pointed out cytotoxic features of some metal nanoparticles.~\cite{verma2008surface, leroueil2008wide} Several studies suggested that metal nanoparticles interact with cell membranes in a complex way,~\cite{roiter2008interaction, chen2010differential, verma2010effect, lin2010penetration} governed by electrochemical potentials and ion distributions around the membrane and a nanoparticle.
These achievements emphasize the importance of a deeper theoretical understanding of the interface between a nanoparticle and the solvent, which acts as a determining factor for many properties of the nanoparticle and its complexes in aqueous environments.~\cite{heikkila2012atomistic}

On a simplified level of theoretical description, a basic elucidation involves an implicit-solvent  treatment of the electrostatic double-layer problem, adopted from the well-established framework of colloidal science. Now of course, the high dielectric interior inverses the role of the image charges as compared to the case of low-dielectric colloidal particles, which thus become attractive and can trigger completely new physics. The attraction between the metal nanoparticle surface and ions can lead to their accumulation and adsorption and thus to a build-up of an electric double layer surrounding the particle, which crucially impacts the colloidal stability~\cite{mafune2000structure, roucoux2002reduced, kim2005control, perera2016counterion, javidpour2018pre} and interactions with other molecules. 
This phenomenon is also of great importance in the catalysis by metal nanoparticles in liquid phase.~\cite{astruc, catalysis} The reaction rates of surface-catalyzed bimolecular reactions depend on the  concentration at the nanoparticle surface of both reactants,~\cite{bimolecular, roy2017revealing} which typically have  asymmetric properties (charge, specific adsorption,~\cite{catalysis} etc.).

In this work, we employ theoretical approaches established in the colloidal electrostatics framework, and apply them to less investigated systems of neutral polarizable nanoparticles in different electrolytes.
We corroborate the theoretical outcomes by Monte Carlo~(MC) simulations, which enable us to assess their regimes of applicability. We show how different kinds of asymmetries between ions result in non-trivial phenomena, such as charge separation and a build-up of net electrostatic potential and  effective surface charge.


\section{Model and Methods}

\begin{figure}\begin{center}
\begin{minipage}[b]{0.3\textwidth}\begin{center}
\includegraphics[width=\textwidth]{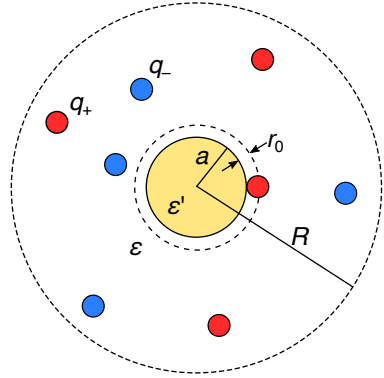}
\end{center}\end{minipage}
\caption{Schematic description of a neutral polarizable nanoparticle of radius $a$ (yellow circle) immersed in an electrolyte solution of cations and anions (red and blue circles). The ions have a radius $r_0$, which specifies the closest-approach distance to other ions and to the nanoparticle.  The solvent is treated as a background continuum with the relative permittivity $\epsilon$ that is much smaller than the permitivity $\epsilon'$ of the nanoparticle. In the MC simulations, the whole system is enclosed in a spherical simulation box of radius $R$ with a reflecting boundary condition.}
\label{fig:model}
\end{center}\end{figure}

We consider a metal nanoparticle as a neutral sphere with a radius $a$ and a relative permittivity $\epsilon'$ that is much larger than the permittivity of the surrounding electrolyte solution $\epsilon$. 
The electrolyte comprises a mixture of cations with valency $q_+$ and anions with valency $q_-$ with bulk concentrations $n_0^{(+)}$ and $n_0^{(-)}$, respectively, by which the electroneutrality condition, $q_+n_0^{(+)}+q_-n_0^{(-)}=0$, has to be fulfilled.
We express all distances by the Bjerrum length at ambient temperature, defined as $\lB =e_0^2/(4\pi\varepsilon\varepsilon_0 \kB T)$ (in water at room temperature, the value is $\lB=0.72~$nm), where $\kB$ is the Boltzmann constant and $T$ the absolute temperature.
The ions are treated as spherical charges with the radius $r_0=0.2\lambda_\trm B$, which specifies the closest approach to other ions as well as to the nanoparticle surface (see \Fig~\ref{fig:model}).

The presence of the dielectric inhomogeneity across the boundary of the sphere influences the electrostatic potential, which can be thus described by the Green's function connecting two points $\Av r$ and $\Av r'$ outside the sphere as
\begin{equation}
u(\Av r, \Av r')=u_0(\Av r, \Av r')+u_\trm {im}(\Av r, \Av r')
\label{eq:u_image}
\end{equation}
Here, $u_0$ is the direct standard Coulomb kernel in the absence of the dielectric inhomogeneities,
\begin{equation}
u_0(\Av r, \Av r')=\frac{1}{4\pi\epsilon\epsilon_0|\Av r-\Av r'|}
\label{eq:u0_image}
\end{equation}
and $u_\trm{im}$ is the ``image correction'' term (sometimes referred to also as the ``reaction field'') due to the dielectric jump, just as in the case of a planar discontinuity. It turns out that for spherical dielectrics in the limit $\epsilon'\gg\epsilon$ there is an elegant analytical solution for the electrostatic potential, which is obtained with the help of ``image charges''.~\cite{friedman1975image, Linse1} Namely, the electrostatic potential of a point charge $q$ at distance $r'$ from the center of the sphere is the same as if there were two additional ``image'' charges instead of the sphere: one in the center of the sphere with the charge $q (a/r')$ and the other one with the charge $-q (a/r')$ dislocated by $a^2/r'$ from the center on the line towards the real charge,
\begin{equation}
u_\trm{im}(\Av r, \Av r')=\frac{a/r'}{4\pi \varepsilon\varepsilon_0 \left|\Av r \right|}-
\frac{a/r'}{4\pi \varepsilon\varepsilon_0 \bigl|\Av r-\tfrac{a^2}{r'^2}\Av r'\bigr|}
\label{eq:uim_image}
\end{equation}
With this exact Green's function at hand, we now turn to investigate the behavior of ions in the proximity of neutral metal nanoparticles in terms of analytical theories as well as MC simulations.


\subsection{Theoretical approaches}
The most common theoretical framework for treatment of electrostatics problems is the Poisson--Boltzmann~(PB) equation, based on mean-field premises.~\cite{holm2012electrostatic, dean2014electrostatics} As such, it {\it cannot} describe any image-charge effects on its own. Therefore, for our setting, the PB equation yields a trivial result of non-perturbed ionic distributions around the nanoparticle. In order to account for the polarizability effects, we follow the original ideas of Onsager and Samaras~\cite{onsagersamaras}: We first calculate the {\it self-energy} (\ie, the potential of mean force) of an ion near the dielectric boundary and then combine it with the PB equation.

\subsubsection{Onsager--Samaras self-energy}
The simplest theoretical treatment to calculate the self-energy of the above introduced model would be to ignore interactions between ions and considering only the image-charge attraction of an ion with the nanoparticle as dictated by \Eq~(\ref{eq:uim_image}). The self-energy of the ion in this approach is given simply by the interaction potential of the ion with its own images, that is
$(1/2) e_0^2 u_\trm{im}(r,r)$. But due to the screening action of the ``ionic atmosphere'' caused by surrounding ions, the image force is considerable only within distances of the order of the Debye length from the surface, defined in terms of the screening coefficient $\kappa$ (\ie, the inverse Debye length) as
\begin{equation}
\kappa^2=4\pi \lB \sum_i q_i^2 n_0^{(i)}
\end{equation}
where the sum runs over all ion species.
In order to heal the impairments stemming from the surrounding ions, Onsager and Samaras~\cite{onsagersamaras} proposed a ``screening coefficient'' to the image charge in the form of $\exp(-2\kappa z)$, where $z$ is the distance of the charge from the dielectric plane. 
\chgA{
The factor of 2 in the exponent arises because the total ``action--reaction'' screening distance from the ion to the surface and back to the ion is $2z$.
Besides, the distance $2z$ corresponds also to the distance between the ion and its virtual image charge in the planar geometry.}
Note that Onsager and Samaras originally proposed the correction for the planar geometry as a first approximation in order to simplify the laborious calculations by Wagner~\cite{wagner} who used a spatially varying screening length.
Their primary aim was to compute the excess surface tension of electrolyte solutions by integrating the Gibbs adsorption equation.~\cite{onsagersamaras}

\chgA{In our first two approaches, we adopt the screening coefficient of Onsager and Samaras to derive an {\it approximate} image self-energy of a monovalent ion near a metal sphere. Yet, in the spherical geometry, we have at least two possibilities of adapting the screening distance $2z$.  In the first approach, we assume twice the distance between the ion and the sphere surface, $2(r-a)$, which gives the self-energy}
\begin{equation}
w_0^\trm{OS}(r)=(1/2) e_0^2\, u_\trm{im}(r,r) \rme^{-2\kappa (r-a)}
\end{equation}
In the combination with the spherical non-screened image interaction $u_\trm{im}$ given by \Eq~(\ref{eq:uim_image}), this yields a simple analytical expression (rescaled by the thermal energy $\beta^{-1}=\kB T$), which we will refer to as ``Onsager--Samaras''~(OS) image-charge interaction
\begin{equation}
\beta w_0^\trm{OS}(r)=-\frac{\lB a^3}{2 r^2 (r^2-a^2)}\rme^{-2\kappa (r-a)}
\label{eq:OS}
\end{equation}
This interaction can be also seen as the adsorption potential of a monovalent ion to the metal nanoparticle.

\chgA{
In our second approach, we consider the screening distance as the separation between the ion and its images. In this case, each of the two induced image charges  is screened by its own screening coefficient, which is $\exp(-\kappa r)$ for the first and $\exp[-\kappa r(r-a^2/r)]$ for the second image term in \Eq~(\ref{eq:uim_image}).
The resulting Onsager--Samaras* (OS*) expression of this approach (which we denote with an asterisk) is then
\begin{equation}
\beta w_0^\trm{OS*}(r)=-\frac 12 \lB a\left[\frac{\exp(\kappa a^2/r)}{r^2-a^2}-\frac{1}{r^2}\right] \rme^{-\kappa r}
\label{eq:OSaster}
\end{equation}
}

\chgA{Note that both expressions, OS and OS*, have not been} self-consistently derived but obtained by \chgA{an {\it ad-hoc} ``stitching'' together} the effects of dielectric discontinuity and ionic screening, and \chgA{are} therefore not exact. 
\chgA{Consequently, it is also not {\it a priori} clear, which of the two approaches yields more accurate results.}

\subsubsection{Debye--H\"uckel self-energy}
In our \chgA{third} approach, we base the image self-energy on the exact Green's function $u^\trm{DH}(\Av r, \Av r')$ of the Debye--H\"uckel~(DH) equation in the presence of a metal sphere,~\cite{spherical2, javidpour2018pre} viz.
\begin{equation}
\nabla^2u^\trm{DH}-\kappa^2u^\trm{DH}=-\frac{1}{\varepsilon\varepsilon_0}\delta(\Av r-\Av r')
\label{eq:DH}
\end{equation}
The Green's function simultaneously accounts for dielectric and screening discontinuities at the surface of the metal sphere. 
The derivation details are provided in Appendix. The final result for the ``DH image self-energy'' of a monovalent ion reads
\begin{equation}
\beta w_0^\trm{DH}(r)=-\frac{\lB \kappa}{\pi}\sum_{l=0}^\infty(2l+1) \eta_l(\kappa a) {k_l}^2(\kappa r)
\label{eq:DHw0}
\end{equation}
with 
\begin{equation}
\eta_l(x)=\left\{
	\begin{array}{ll}
	i_0'(x)/k_0'(x)&\textrm{for $l=0$}\\
	i_l(x)/k_l(x)&\textrm{for $l\ge1$}\\
	\end{array} \right.
\end{equation}
Here, the primes denote derivatives of the spherical modified Bessel functions of the first and second kind, which are defined as
\begin{equation}
i_l(x)=\sqrt{\frac{\pi}{2x}}\,I_{l+1/2}(x)\quad\textrm{and}\quad
k_l(x)=\sqrt{\frac{\pi}{2x}}\,K_{l+1/2}(x)
\label{eq:bessel}
\end{equation}
where $I_{l+1/2}(x)$ and $K_{l+1/2}(x)$ are the conventional modified Bessel functions of the first and second kind, respectively.

In the limit of infinitely large radius $a\to\infty$ (\ie, planar metal wall), both \Eq~(\ref{eq:OS}) and~(\ref{eq:DHw0}) simplify to 
\begin{equation}
\beta w_0(z)\simeq -\frac{\lB}{4z}\,\rme^{-2\kappa z}
\label{eq:w0-planar}
\end{equation}
where $z$ is the distance of the ion from the wall. Exactly the same expression but with the opposite sign applies in the case of an ion near a planar wall with much lower dielectric interior than the electrolyte solution ($\varepsilon'\ll\varepsilon$).~\cite{kanduc2011dressed}

\subsubsection{Boltzmann distribution}

In cases when cations and anions have symmetric properties, no electrostatic potential is generated, and their distribution around the nanoparticle is solely governed by the image self-energy. In a thermodynamic equilibrium, we therefore expect the ion densities to follow the Boltzmann distribution. Using the OS [\Eq~(\ref{eq:OS})] and OS* [\Eq~(\ref{eq:OSaster})] self-energies, this leads respectively to 
\begin{equation}
n^{(i)}(r)=n_0^{(i)} \exp\bigl[-\beta q_i^2 w_0^\trm{OS}(r)\bigr]\quad\trm{(B--OS)}
\label{eq:B-OS}
\end{equation}
\chgA{and
\begin{equation}
n^{(i)}(r)=n_0^{(i)} \exp\bigl[-\beta q_i^2 w_0^\trm{OS*}(r)\bigr]\quad\trm{(B--OS*)}
\label{eq:B-OSaster}
\end{equation}
which we will term as ``Boltzmann--Onsager--Samaras'' approximations (B--OS and B--OS*, respectively).}
Similarly, using the DH form~(\ref{eq:DHw0}), gives us
\begin{equation}
n^{(i)}(r)=n_0^{(i)} \exp\bigl[-\beta q_i^2 w_0^\trm{DH}(r)\bigr]\quad\trm{(B--DH)}
\label{eq:B-DH}
\end{equation}
which we term the ``Boltzmann--Debye--H\"uckel''~(B--DH) approximation.
Here, $q_i$ is the valency and $n_0^{(i)}$ the bulk concentration of species $i$.



\subsubsection{Modified Poisson--Boltzmann}
If cations and anions redistribute dissimilarly around the nanoparticle, the resulting charge separation can lead to a net electrostatic potential, and thus \Eqs~(\ref{eq:B-OS}) and  (\ref{eq:B-DH}) become inaccurate.
As already mentioned, the standard PB equation, which relates electrostatic potential and charge distributions, lacks the image self-energy term.
A simple heuristic ``remedy'' to account for the image effects is to insert by hand the self-energy correction into the Boltzmann factor, thus leading to a modified Poisson--Boltzmann equation~\cite{corry2003dielectric, onuki2006ginzburg} in the form
\begin{equation}
\nabla^2\phi(\Av r) = -\frac {1}{\varepsilon\varepsilon_0}\sum_{i}q_i n_0^{(i)}\exp\bigl[-\beta q_i\phi(\Av r)-\beta q_i^2 w_0^\trm{DH}(\Av r)\bigr]
\label{eq:mPB}
\end{equation}
Here, the summation runs over all ion species $i$.
The self-energy term is in principle given either by \chgA{\Eq~(\ref{eq:OS}), (\ref{eq:OSaster}), or (\ref{eq:DHw0})}. In our analyses, however, we will limit ourselves only to the DH-based form, \Eq~(\ref{eq:DHw0}).
Once the potential $\phi(r)$ is known, the ion densities can be evaluated as
\begin{equation}
n^{(i)}(\Av r)=n_0^{(i)}\exp\bigl[-\beta q_i\phi(\Av r)-\beta q_i^2 w_0^\trm{DH}(\Av r)\bigr]\quad\trm{(PB--DH)}
\label{eq:PB-DH}
\end{equation}
which we term as the ``Poisson--Boltzmann--Debye--H\"uckel'' (PB--DH) approach in this paper.
 In the case of a symmetry between cations and anions, the electrostatic potential vanishes, $\phi=0$, and \Eq~(\ref{eq:PB-DH}) reduces to \Eq~(\ref{eq:B-DH}).


Note again, that the obtained expressions, \Eqs~(\ref{eq:B-OS}--\ref{eq:PB-DH}), cannot be considered as mean-field results, because they do not follow from the PB equation.
Even though many studies~\cite{diamant1996kinetics, dean2004field, levin2009ions, schwierz2010reversed, markovich2015surface} generalized the seminal work of Onsager and Samaras, it nevertheless remains widely misinterpreted what is the actual theoretical framework of  their approach.
In fact, these results extend beyond the mean-field level and can be deduced from the thermodynamic fluctuations of the instantaneous electric fields around the PB solution.~\cite{podgornik1988inhomogeneous}
Alternatively, the PB--DH equation can be derived from a self-consistent variational analysis~\cite{wang,weaklycharged,inhomogeneous2,fieldtheory} by setting {\it by hand} the screening length $\kappa$ to be location independent.


\subsection{Monte Carlo simulations}
In order to provide ``exact'' solutions to the introduced model, we perform MC simulations in the canonical NVT ensemble using the standard Metropolis algorithm.~\cite{metropolis1953equation}
The system with mobile ions is enclosed in a spherical simulation box with an outer radius $R$, containing $N_+$ cations and $N_-$ anions with valencies $q_+$ and $q_-$, respectively, such that their amounts fulfill the electroneutrality condition $N_+q_++N_-q_-=0$, see \Fig~\ref{fig:model}.
A reflecting boundary condition is applied to the external box boundary.
As opposed to periodic boundary conditions, this treatment significantly simplifies the implementation and increases the performance of the simulations (as no Ewald summation is needed), whereas it distorts ionic distributions near the outer boundary. 
In all simulations, the radius of the spherical box is set to $R=17\,\lambda_\trm{B}$, which is significantly larger than the largest Debye length of $\kappa^{-1}\approx 9\,\lambda_\trm{B}$ in the study. 
This guarantees that the outer boundary does not impact the ionic behavior near the nanoparticle. 

\section{Results and discussions}

\subsection{Symmetric case}

\begin{figure*}[t]\begin{center}
\begin{minipage}[b]{0.6\textwidth}\begin{center}
\includegraphics[width=\textwidth]{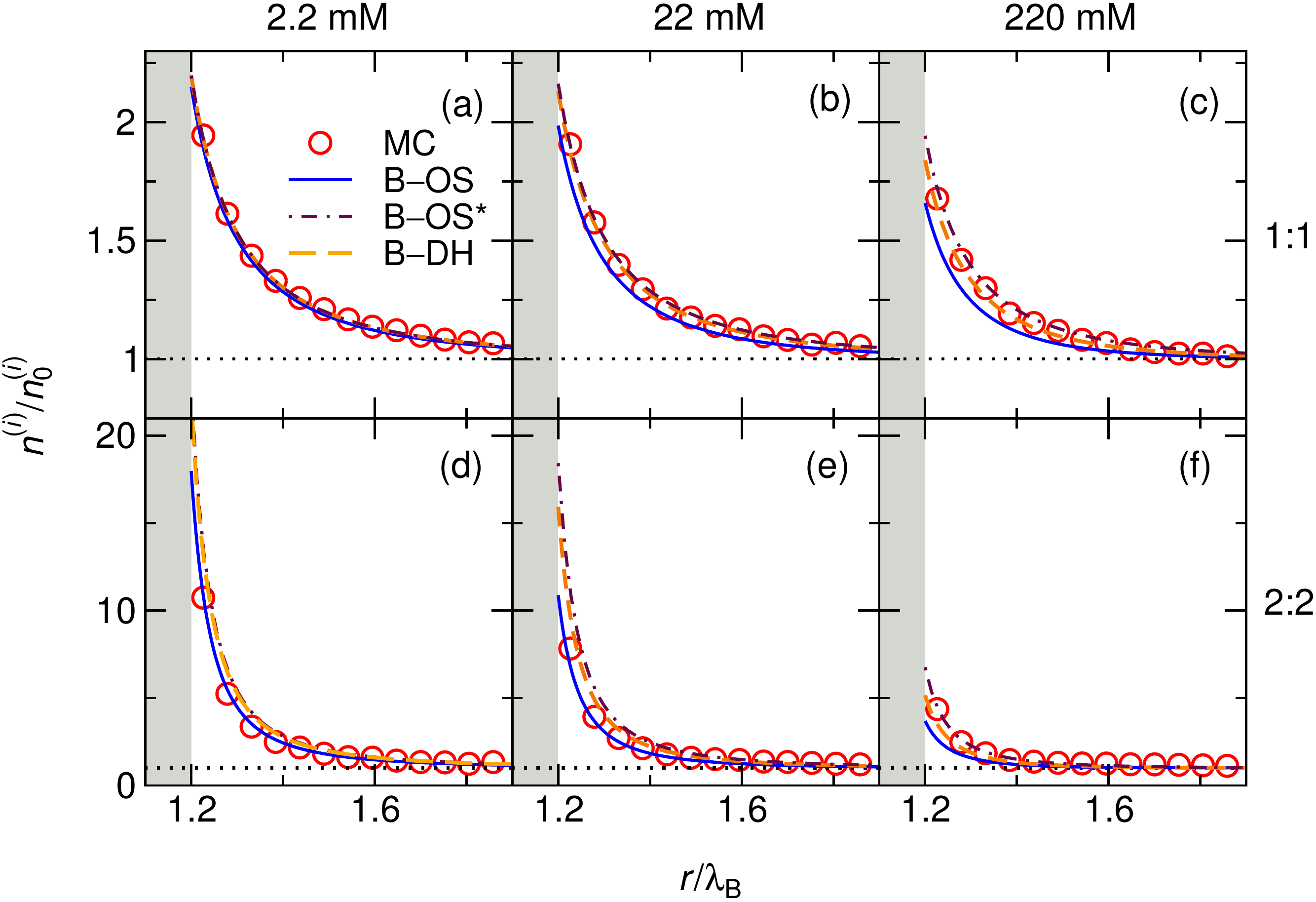} 
\end{center}\end{minipage}
\caption{Ionic densities for symmetric 1:1 (top panels) and 2:2 (bottom panels) electrolytes of concentrations 2.2~mM (left), 22~mM (middle), and 220~mM (right) at the metal nanoparticle with the radius $a=\lB$. The shaded areas denote an inaccessible region to ions, $r<a+r_0$, where $r_0=0.2\lB$ is the radius of the ions.}
\label{fig:symmetric}
\end{center}\end{figure*}

We start our theoretical analysis by first considering symmetric electrolytes ($q_+=-q_-$). 
In \Fig~\ref{fig:symmetric} we plot ion profiles for 1:1~(top) and 2:2~(bottom) cases at three different bulk concentrations (from left to right: 2.2, 22, and 220~mM) at a nanoparticle of size $a=\lB$ as predicted by \chgA{all three theoretical approaches \Eqs~(\ref{eq:B-OS}--\ref{eq:B-DH})} and MC simulations. Since in this case the cations and anions have symmetric properties, their density distributions are equivalent.
As seen from the plots, ions are considerably attracted to the metal nanoparticle surface due to attractive image charge interactions. The density peaks right at the surface vicinity (at $r=a+r_0$) and is therefore highly sensitive to the minimum approach distance of an ion to the surface, or to be more precise, to the dielectric boundary, in our model determined by the ion radius $r_0$.
From the OS equation (\ref{eq:OS}), it can be easily appraised that the attractive adsorption energy at the surface (\ie, at $r=a+r_0$), scales as $w_0\sim -1/r_0$, thus making it very sensitive to the choice of $r_0$. Nevertheless, we will keep the value fixed at $r_0=0.2\lB$ for all further results in this study.

For very low ion concentrations (2.2~mM), the screening length is considerably large ($\kappa^{-1}=9\lB$ for the monovalent and $4\lB$ for the divalent case), such that the interaction near the surface is predominantly governed by the unscreened part of the image charge interaction. Moreover, in the limit of vanishing salt concentration, \chgA{all three} theories become equivalent and exact.
In the cases shown in the figure, \chgA{all} the theories agree very well even for salt concentrations up to 220~mM.


\begin{figure*}[t]\begin{center}
\begin{minipage}[b]{0.69\textwidth}\begin{center}
\includegraphics[width=\textwidth]{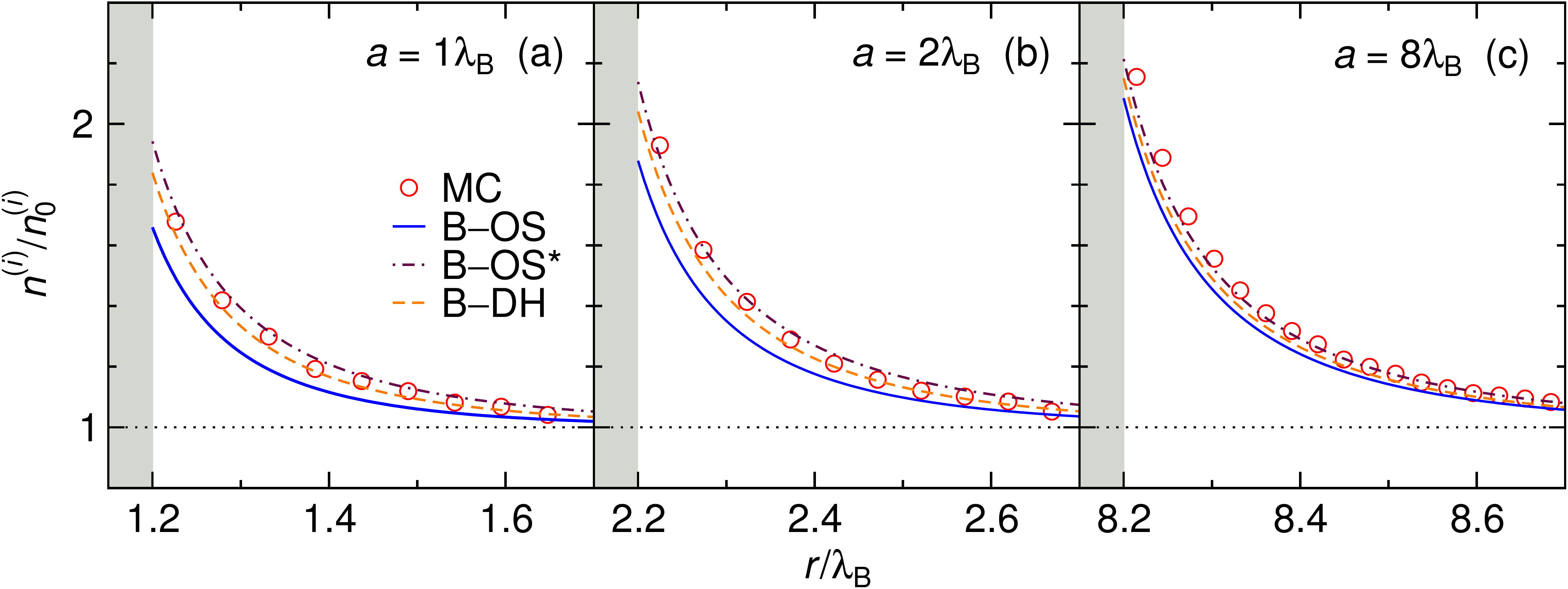} 
\end{center}\end{minipage}\hspace{2ex}
\begin{minipage}[b]{0.255\textwidth}\begin{center}
\includegraphics[width=\textwidth]{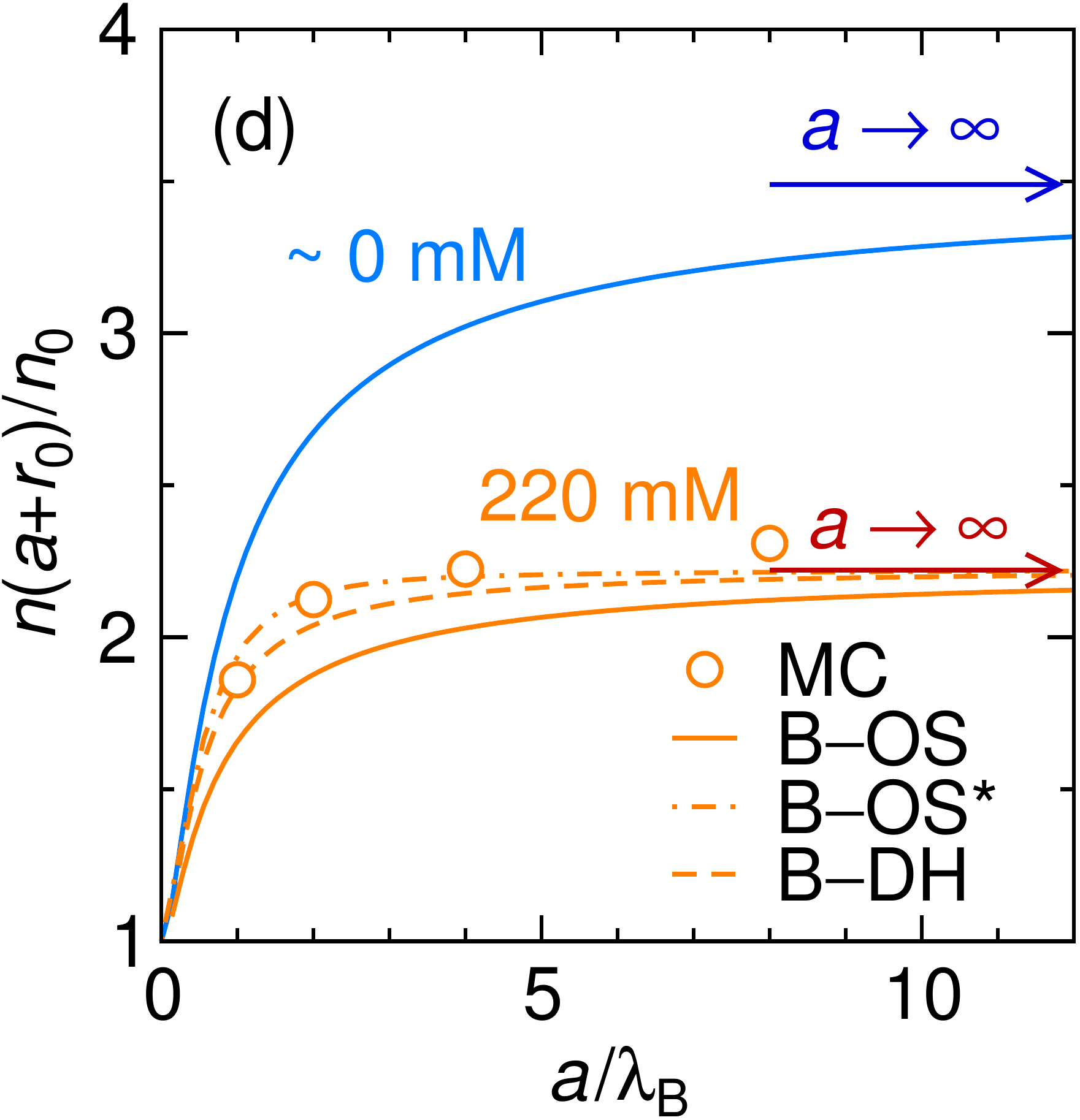} 
\end{center}\end{minipage}
\caption{The influence of the nanoparticle size. 
(a--c) Ion densities at nanoparticles of different radii $a$ in a 1:1 electrolyte with the bulk concentration of 220~mM. The shaded areas indicate the excluded region.
(d)~Ion density at the nanoparticle surface, $r=a+r_0$, as a function of the its radius $a$ for vanishing salt concentration, $\sim $~0~mM, and for 220~mM. 
In the case of vanishing concentration \chgA{the} theories become exact (blue solid line). The arrows indicate the theoretical predictions at a flat interface (\chgA{all three} theories become equivalent).
The ion size is set to $r_0=0.2\lB$.}
\label{fig:size}
\end{center}\end{figure*}

The size of the nanoparticle is another important parameter that determines the strength of the image attraction.
To demonstrate this effect, we plot in \Fig~\ref{fig:size} the density profiles for a monovalent 1:1 electrolyte at 220 mM for different radii $a$ of the nanoparticle.
 With an increasing size, the densities at the surface get higher.
Larger metal nanoparticles have namely higher polarizability, thus attracting the ions more efficiently. From the plot it can also be observed that \chgA{all three} theories are becoming equivalent as the particle size increases. 
\Figure~\ref{fig:size}d shows a normalized ion density at the nanoparticle surface (\ie, at $r=a+r_0$) as a function of its radius $a$. In the limiting case of vanishing particle ($a\to 0$), clearly, the polarizability vanishes and the density becomes bulk-like. The density increases with the radius and saturates at the limit of a planar wall (indicated by arrows), where the interaction is given by \Eq~(\ref{eq:w0-planar}).
In the limit of vanishing ionic strength (blue solid line), both theories become exact, since in that case the ion--ion interactions become rare and negligible.
For higher concentrations (220~mM), the particle of size of $a\sim 2\lB$ already nearly reaches the planar-wall limit. Here, it can also be noted that \chgA{the theories (except B--OS* in the case of small particles)} tend to slightly underestimate the densities near the surface compared with MC simulations. 
This can be of course attributed to several effects. One of them might be the absence of screening in the ion-free layer of the width $r_0$ around the surface, which has been for instance discussed by Levin and Mena.~\cite{levinmena}
 
\chgA{Coming to the question of which of the three theoretical approaches is the most accurate: It is of course expected that B--DH should predict more accurate results than either B--OS or B--OS*, because it properly takes into account the spherical geometry of the problem on the DH level. 
As can be seen from \Figs~\ref{fig:symmetric} and~\ref{fig:size}, the results of both approximate theories are very close to the results of B--DH.
Interestingly, B--OS seems to consistently yield a bit lower densities than B--DH, meaning that it underestimates the overall attraction of the ion to the metal sphere.
On the contrary, B--OS* predicts consistently slightly larger results than B--DH.
It seems that for small spheres, the B--OS* performs slightly better than B--OS. However, this cannot be claimed for larger spheres, as shown in \Fig~\ref{fig:size}c, where both B--OS and B--OS* are approximately equally off, yet in opposite directions.
However, the advantage of the approximate OS and OS* expressions is their much simpler mathematical form than B--DH.
}

\subsection{Asymmetric case: specific adsorption}

Continuum theoretical descriptions based on the dielectric approximation generally treat ions as equivalent point charges and neglect the nonelectrostatic interactions between ions and the particle surface, which occur in realistic systems.~\cite{collins2012continuum} The origin of these ion-specific interactions is still the subject of vivid debate, but in recent years, it has become well established that they are mainly influenced by three parameters: ion--surface, ion--water, as well as water--surface interactions, namely hydrophilicity and hydrophobicity.~\cite{manciu2003specific, parsons2011hofmeister, dos2013surface, schwierz2016reversed}
Different ions are expected to bind to nanoparticle surfaces with different affinities, which typically follow the Hofmeister series.~\cite{pfeiffer2014interaction, merk2014situ, schwierz2016reversed}
Binding of ions to the nanoparticle significantly influences their surface charge and the surface potential, which are crucial for the stability of colloidal suspensions based on electrostatic repulsion.~\cite{kim2005control, merk2014situ, perera2016counterion}


\begin{figure}[h]\begin{center}
\begin{minipage}[b]{0.25\textwidth}\begin{center}
\includegraphics[width=\textwidth]{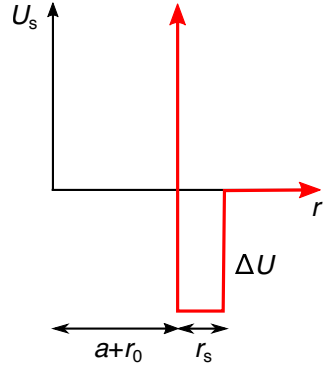} 
\end{center}\end{minipage}
\caption{Additional specific adsorption potential for cations.}
\label{fig:pot_specific}
\end{center}\end{figure}

On the continuum-level description, the specific effects can be phenomenologically incorporated via various approximate approaches. In the simplest approximation, the specifically adsorbed ions in the Stern layer close to the surface can be, for instance, treated as a fixed pre-determined surface charge, which is a concept adopted in many theoretical approaches. The main shortcoming of this approximation is that it neglects the dependence of the adsorbed amount of ions on the bulk concentration. Furthermore, it also neglects the influence of surface polarizability and ion correlations.
Another approach, which we will adopt here, is to assume an additional attractive potential $U_\trm{s}(r)$ between the ions and the nanoparticle. For simplicity, we use a square-well potential of depth $\Delta U=-2\kB T$ and the range of $r_\textrm{s}=0.3\lB$ from the effective nanoparticle surface, as presented in \Fig~\ref{fig:pot_specific}. In order to introduce an asymmetry in our system, we apply this potential only to cations, while we assume no specific interactions for anions.
  We plug the potential $U_\trm{s}(r)$ into the Boltzmann factor of \Eqs~(\ref{eq:B-DH}) and (\ref{eq:PB-DH}).
This break of the symmetry, assumed in the previous section, has far-reaching consequences as we will see in the following.

\begin{figure*}[t]\begin{center}
\begin{minipage}[b]{0.9\textwidth}\begin{center}
\includegraphics[width=\textwidth]{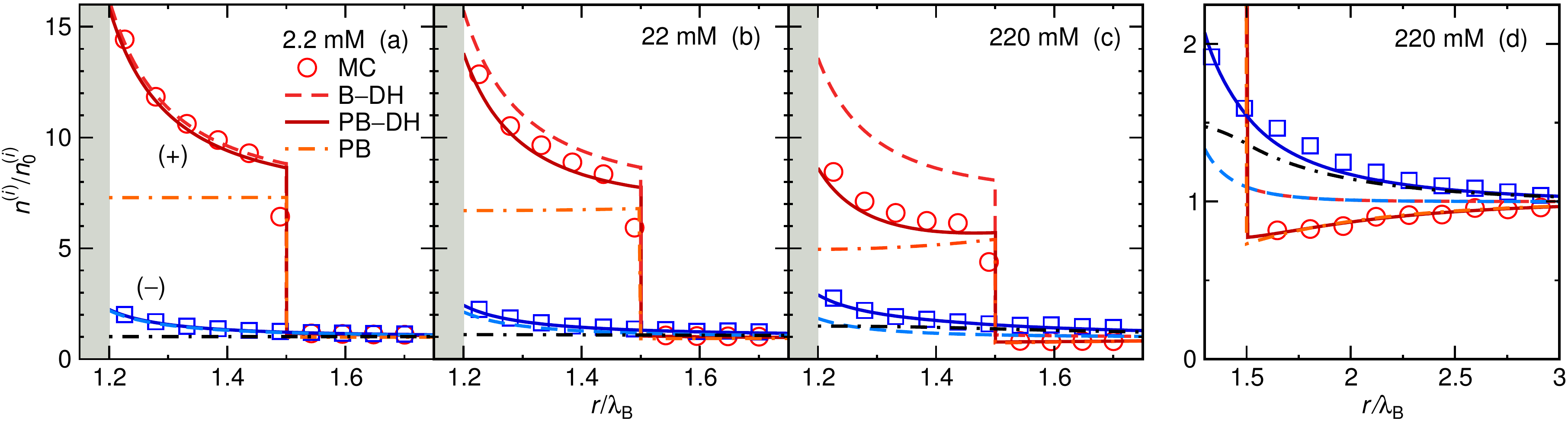} 
\end{center}\end{minipage}
\caption{Ion densities profiles at a nanoparticle of radius $a=\lB$ with the specific adsorption of $\Delta U=-2\kB T$ for cations in (a)~2.2~mM, (b)~22~mM, and (c)~220~mM of 1:1 electrolyte. The red-shaded curves correspond to cation (+) densities and the blue-shaded to  anion (--) densities.
(d)~Far-field region of the case in~(c).}
\label{fig:dens_specific}
\end{center}\end{figure*}

The ion distributions, shown in \Fig~\ref{fig:dens_specific}, now exhibit a distinct accumulation of cations due to the specific adsorption potential.
Notably, the simple Boltzmann-based approach B--DH [\Eq~(\ref{eq:B-DH})] already captures the densities sufficiently well at low concentrations, since the generated electrostatic potential has  negligible influence on ions. 
But as we increase the concentration, the relative cation density $n^{(+)}(r)/n_0^{(+)}$ near the surface starts to decrease and anion density slightly to increase. Namely, the potential generated by the adsorbed cations is hindering further accumulation of cations. This behavior is well captured by PB--DH [\Eq~(\ref{eq:PB-DH})], whereas the simple B--DH starts breaking down.
A crucial difference between PB--DH and B--DH shows up when zooming in to the far-field region [panel (d)] that extends beyond the specific adsorption potential. There, the ion distributions are considerably influenced by the generated potential. As can be noted, the anion density is higher than the cationic, since anions have to compensate the accumulated positive charge at the surface.
An interesting comparison can be made when considering only a PB equation with the specific adsorption but without the image-charge self-energy, 
\begin{equation}
\nabla^2\phi(\Av r) = -\frac {1}{\varepsilon\varepsilon_0}\sum_{i}q_i n_0^{(i)}\exp\bigl[-\beta q_i\phi(\Av r)+\beta U_\trm{s}(\Av r)\bigr].
\label{eq:PB}
\end{equation}
The resulting densities are shown in \Fig~\ref{fig:dens_specific}d by dash--dotted lines and are in the vicinity of the surface expectedly flatter and lower than the other results due to the missing image-charge attraction. Nevertheless, the PB provides reasonable agreement for the ion densities for distances beyond the specific potential.


\begin{figure}[h]\begin{center}
\begin{minipage}[b]{0.233\textwidth}\begin{center}
\includegraphics[width=\textwidth]{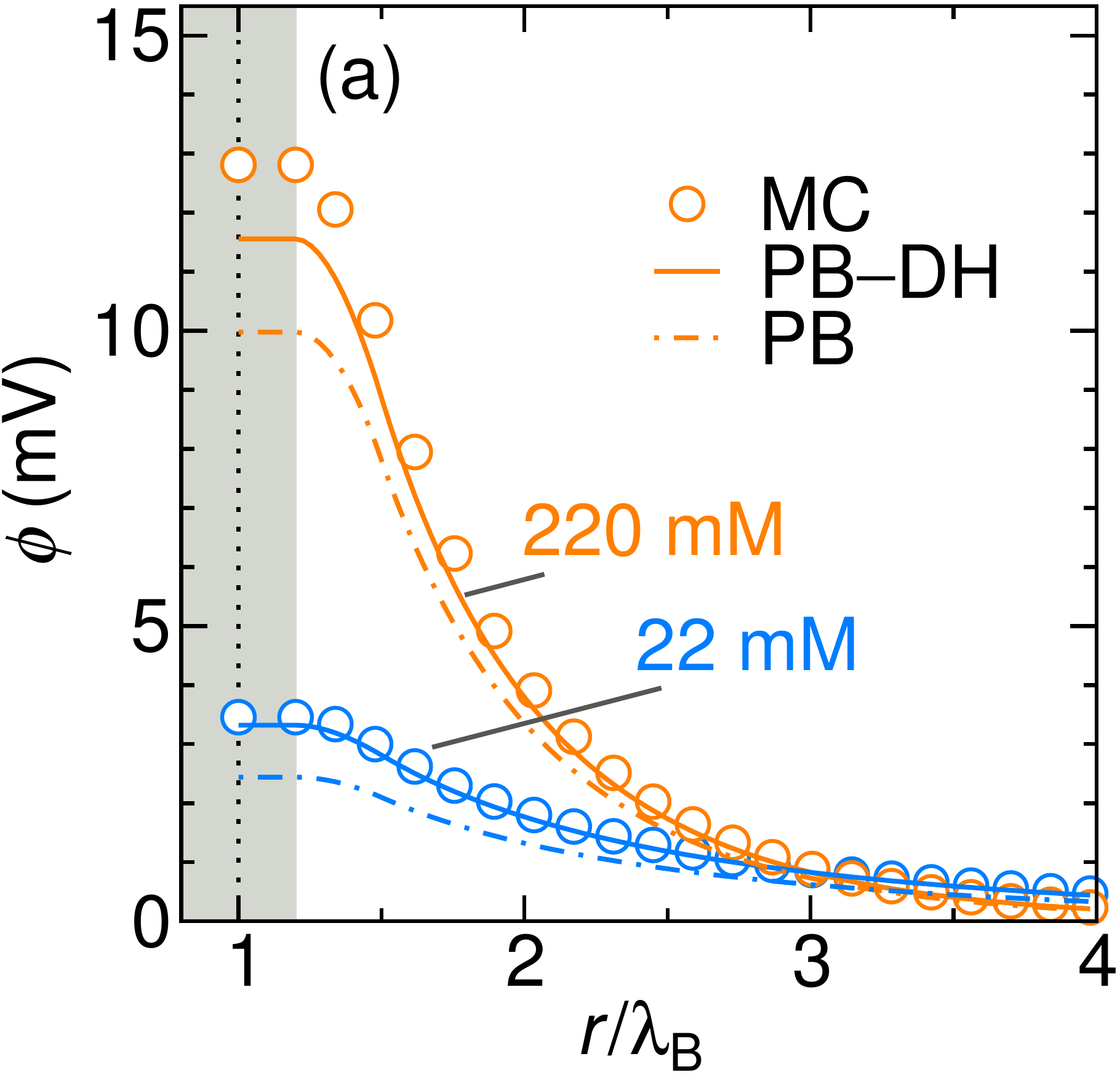} 
\end{center}\end{minipage}
\begin{minipage}[b]{0.235\textwidth}\begin{center}
\includegraphics[width=\textwidth]{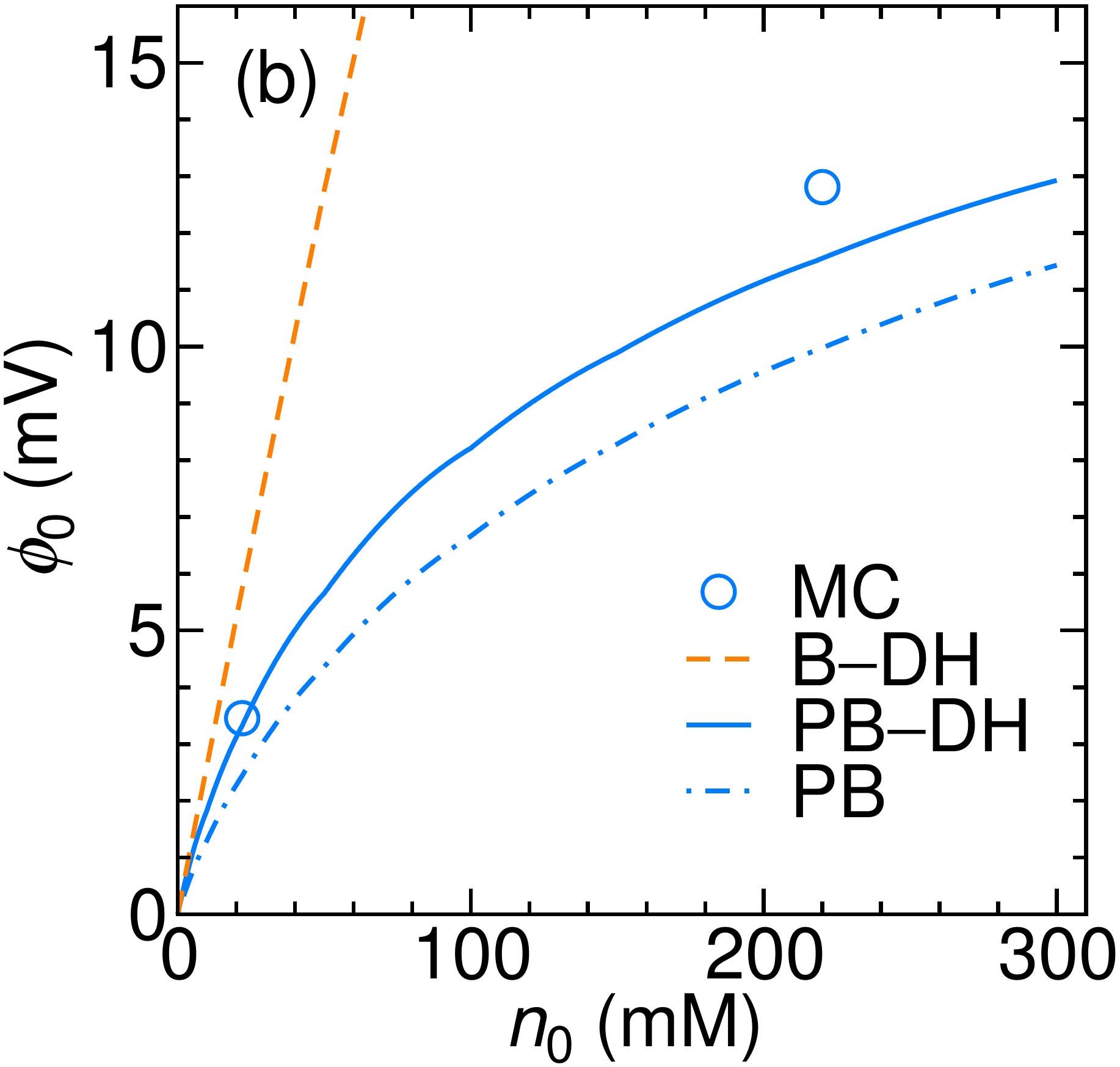} 
\end{center}\end{minipage}\hspace{1ex}
\begin{minipage}[b]{0.237\textwidth}\begin{center}
\includegraphics[width=\textwidth]{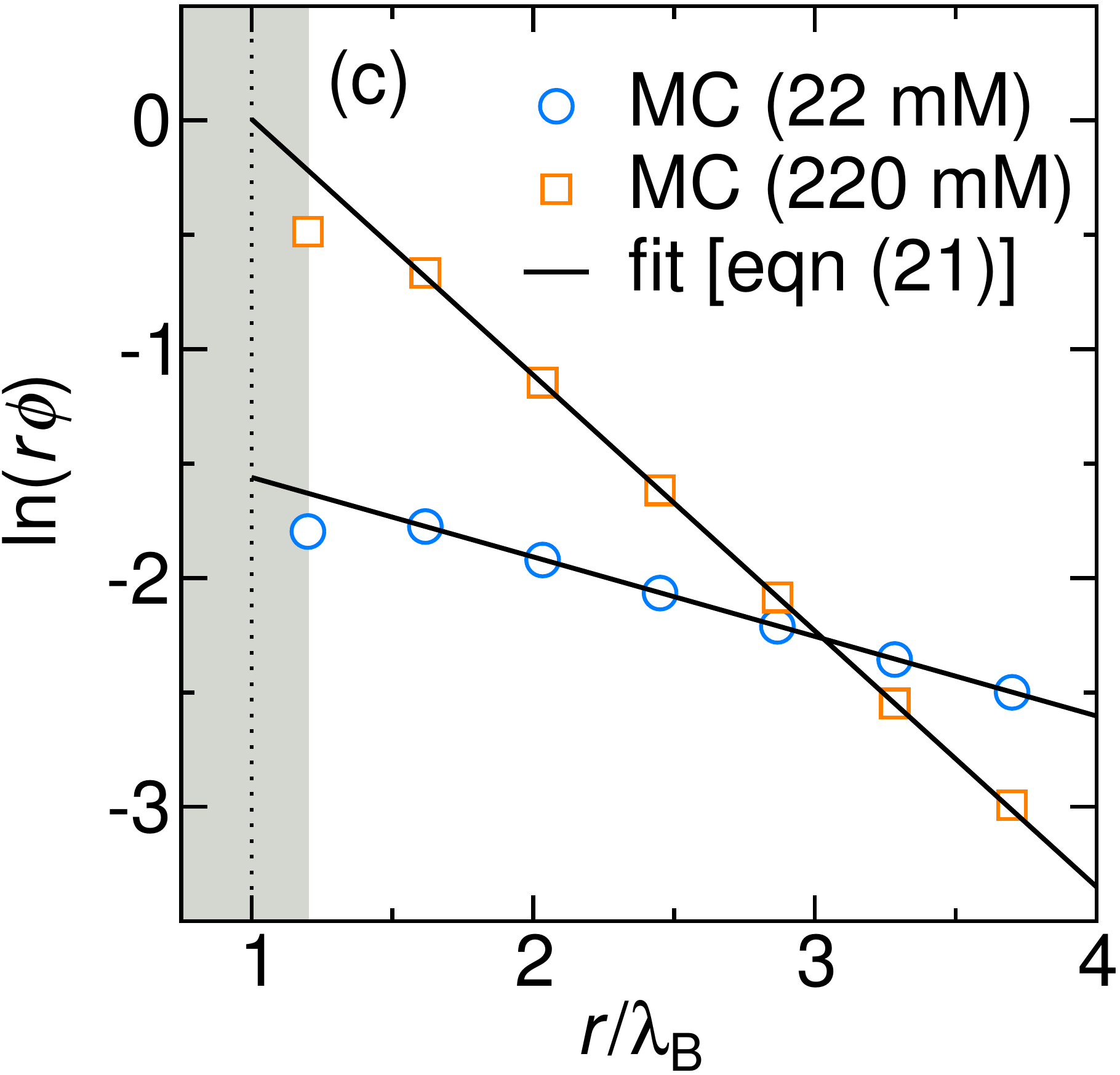} 
\end{center}\end{minipage}
\begin{minipage}[b]{0.23\textwidth}\begin{center}
\includegraphics[width=\textwidth]{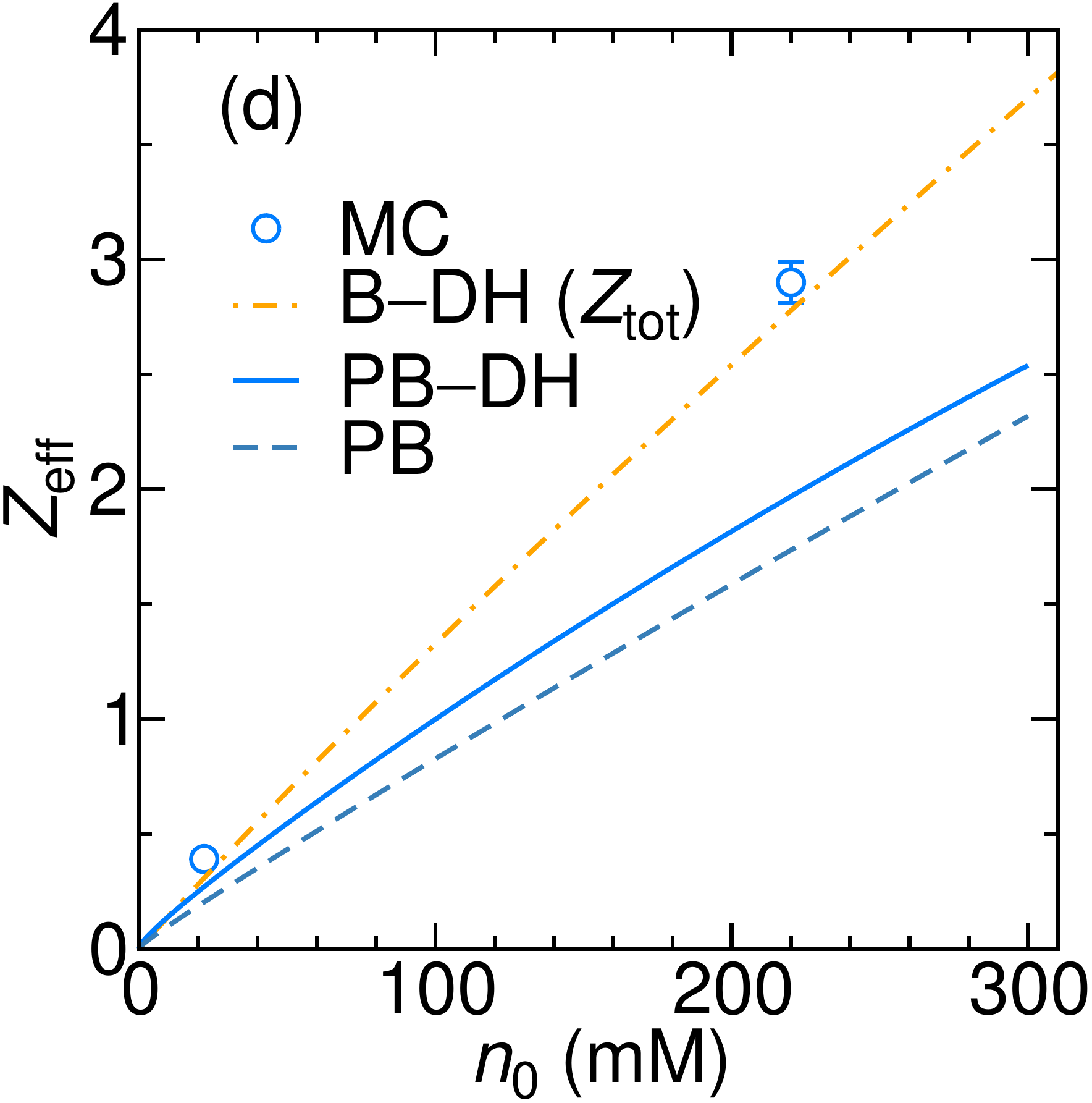} 
\end{center}\end{minipage}
\caption{(a) Generated electrostatic potential at the nanoparticle stemming from the specific adsorption potential as predicted by the PB and PB--DH theories and MC simulations for 22 and 220~mM of 1:1 salt. 
(b)~The corresponding surface potential, $\phi_0=\phi(a)$, as a function salt concentration.
(c)~Linear fit of \Eq~(\ref{eq:DH-Zeff}) to the MC data points for 22 and 220~mM electrolyte concentrations.
(d)~The effective charge of the metal nanoparticle obtained from the fits of \Eq~(\ref{eq:DH-Zeff}) as a function of salt concentration. For the case of B--DH, the total cumulative charge is shown instead.
}
\label{fig:specific}
\end{center}\end{figure}

By integrating the density profiles, we obtain the cumulative charge $Z(r)$ contained within a sphere of radius $r$ around the nanoparticle,
 \begin{equation}
Z(r)=\int_a^r \sum_i q_i n_i(r') \,4\pi r'^2\rmd r'
\label{eq:Z}
\end{equation}
which can be then used to evaluate the electrostatic potential (\eg, in MC simulations) as
\begin{equation}
\phi(r)=\frac{\lambda_\trm{B} \kB T}{e_0}\int_a^{r} \frac{Z(r')}{r'^2}\rmd r'
\label{eq:phi}
\end{equation}
The electrostatic potential generated due to the specific adsorption is shown in \Fig~\ref{fig:specific}a for 22 and 220~mM of a 1:1 electrolyte. The results of PB--DH compare excellent to the MC results. On the other hand, PB that neglects the polarization [\Eq~(\ref{eq:PB})] yields a bit smaller potentials.
It is interesting to examine, how does the surface potential, defined as the electrostatic potential at the nanoparticle surface, $\phi_0=\phi(a)$, evolve with the ionic strength.
As shown in \Fig~\ref{fig:specific}b,  $\phi_0$ first linearly rises with concentration, but the rise is becoming gradually weaker for higher concentrations. This slow-down can be attributed to higher repulsion due to accumulated ions and to more effective screening (larger $\kappa$) at higher concentrations. 
The potential stemming from the B--DH approximation (\ref{eq:B-DH}) can be estimated via the cumulative charge integration, \Eq~(\ref{eq:phi}), as is also done for MC data. Since the B--DH approach neglects the potential, which is especially important for far-field behavior, its predictions are severely off compared with other approaches. The B--DH is therefore in this case only a useful predictor for local ion densities, but it fails for far-field.

As is well established in colloid science, we expect the generated potential $\phi(r)$ to follow a well-known DH law in the far-field,
\begin{equation}
\beta e_0\phi(r)=\frac{\lB Z_\trm{eff}}{1+\kappa a} \frac{\rme{^{-\kappa(r-a)}}}{r}
\label{eq:DH-Zeff}
\end{equation}
where $Z_\trm{eff}$ is the effective charge (normalized by the unit charge $e_0$) of the nanoparticle. Indeed, by fitting \Eq~(\ref{eq:DH-Zeff}) with $Z_\trm{eff}$ and $\kappa$ as fitting parameters to the electrostatic potentials at large distances (shown in \Fig~\ref{fig:specific}c for MC data),  we obtain very good agreement. The effective charge hence arises as a result of charge separation around an otherwise neutral particle. 
\Figure~\ref{fig:specific}d further demonstrates that the effective charge rises almost linearly with the salt concentration. Both PB-based approaches PB--DH and PB predict very good results (comparable to MC) at low salinities, but tend to underestimate (PB slightly more) the values at higher concentrations.
On the other hand, estimating $Z_\trm{eff}$ from the B--DH approach is not possible, since the accumulated charge is effectively not screened by the electrolyte and the evaluated potential in this theory does not follow the DH form of \Eq~(\ref{eq:DH-Zeff}).
Instead,  B--DH predicts a saturation of the cumulative charge $Z(r)$ to a non-zero value, whereas it is realistically expected to vanish at $r\to\infty$, as is the case for the other two theories and MC simulations. Even though, this is due to a deficiency of the B--DH approach, the value reflects the charge accumulation right at the surface, where B--DH performs reasonably well. Therefore, it is interesting to compare the total cumulative charge $Z_\trm{tot}=Z(r\to\infty)$ from B--DH to $Z_\trm{eff}$ from other approaches.
Indeed, in \Fig~\ref{fig:specific}d we see that $Z_\trm{tot}$ compares remarkably to $Z_\trm{eff}$.

\subsection{Asymmetric case: valency}

From specific adsorption we now turn our attention to a different kind of asymmetry, the asymmetry that stems from different ion valencies in an electrolyte.
According to \Eqs~(\ref{eq:B-OS}--\ref{eq:B-DH}), the self-image attraction of an ion to a metal sphere exhibits a square dependence on its charge, $\sim q^2$. Consequently, in cases of  asymmetric electrolytes, such as 2:1 or 3:1, this valency dependence engenders strong differences in adsorption between both ion species.
As opposed to specific adsorption, where the particle polarizablity is only an accompanying effect to asymmetric adsorption and related phenomena, it is the main agent for similar phenomena in the case of asymmteric ion valencies.

\begin{figure}[h]\begin{center}
\begin{minipage}[b]{0.49\textwidth}\begin{center}
\includegraphics[width=\textwidth]{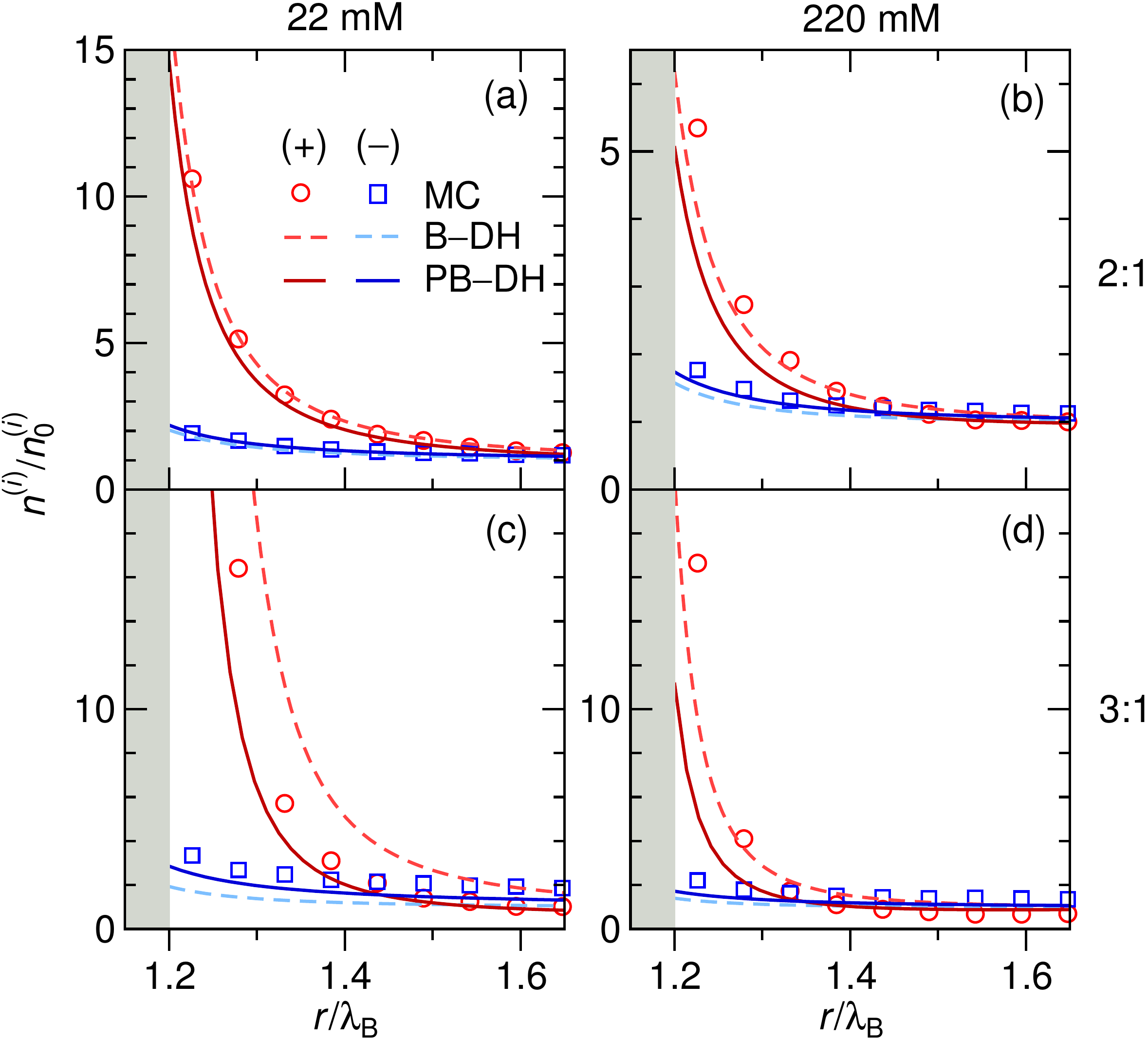} 
\end{center}\end{minipage}
\caption{Normalized ion density profiles for 2:1 and 3:1 electrolytes 
of concentrations 22~mM (left) and 220~mM (right) near a metal nanoparticle of radius $a=\lambda_\trm B$. Cations are considered as the multivalent and anions as the monovalent components.
Theoretical approaches B--DH~[\Eq~(\ref{eq:DHw0})] and PB--DH~[\Eq~(\ref{eq:mPB})] are compare with MC simulation results.
}
\label{fig:dens_valency}
\end{center}\end{figure}

As before, we first look into the ion distributions, which are shown in \Fig~\ref{fig:dens_valency} for asymmetric 2:1 and 3:1 electrolytes, and compare the theoretical approaches B--DH and PB--DH with MC simulations.
As in the case of the ion-specific adsorption, the theories yield better results at low salt concentrations. At higher concentrations, they perform worse due to delicate ion--ion interactions, in particular for higher asymmetry (\ie, 3:1). This theoretical break-down is not unexpected, since multivalent ions are known for significant correlation effects, not accounted for on a mean-field level, a feature that is well established in the double layer literature.~\cite{IV, grosberg2002colloquium, levin2002electrostatic,  boroudjerdi2005statics, matej, outhwaite4, kanduc2017interactions}
As such, \Fig~\ref{fig:dens_valency} demonstrates a dramatic influence of the valency on the local  densities of ions. The relative ionic density at the surface, $n(a+r_0)/n_0$, scales namely as $\sim\exp(\trm{const.}\times q^2)$, which for low ionic strengths leads ``only'' to around 2-fold enrichment of monovalent ions in our system (\Fig~\ref{fig:symmetric}), 16-fold ($\sim 2^4$) of divalent, and an enormous 512-fold ($\sim 2^9$) enrichment of trivalent ions compared to bulk.
This implies high ability of metal particles to take-up multivalent ions from a solution. 
Cases of highly asymmetric electrolytes are very relevant also in catalytic science, where one of well-studied benchmark ``model reactions'' involves the reduction of trivalent hexacyanoferrate~(III) ions by monovalent borohydride ions catalyzed by metal nanoparticles.~\cite{carregal2009colloidal, catalysis} The local density of the reactant at the surface is one of the governing factors that determines the reaction rate.~\cite{bimolecular}

\begin{figure}[h]\begin{center}
\begin{minipage}[b]{0.234\textwidth}\begin{center}
\includegraphics[width=\textwidth]{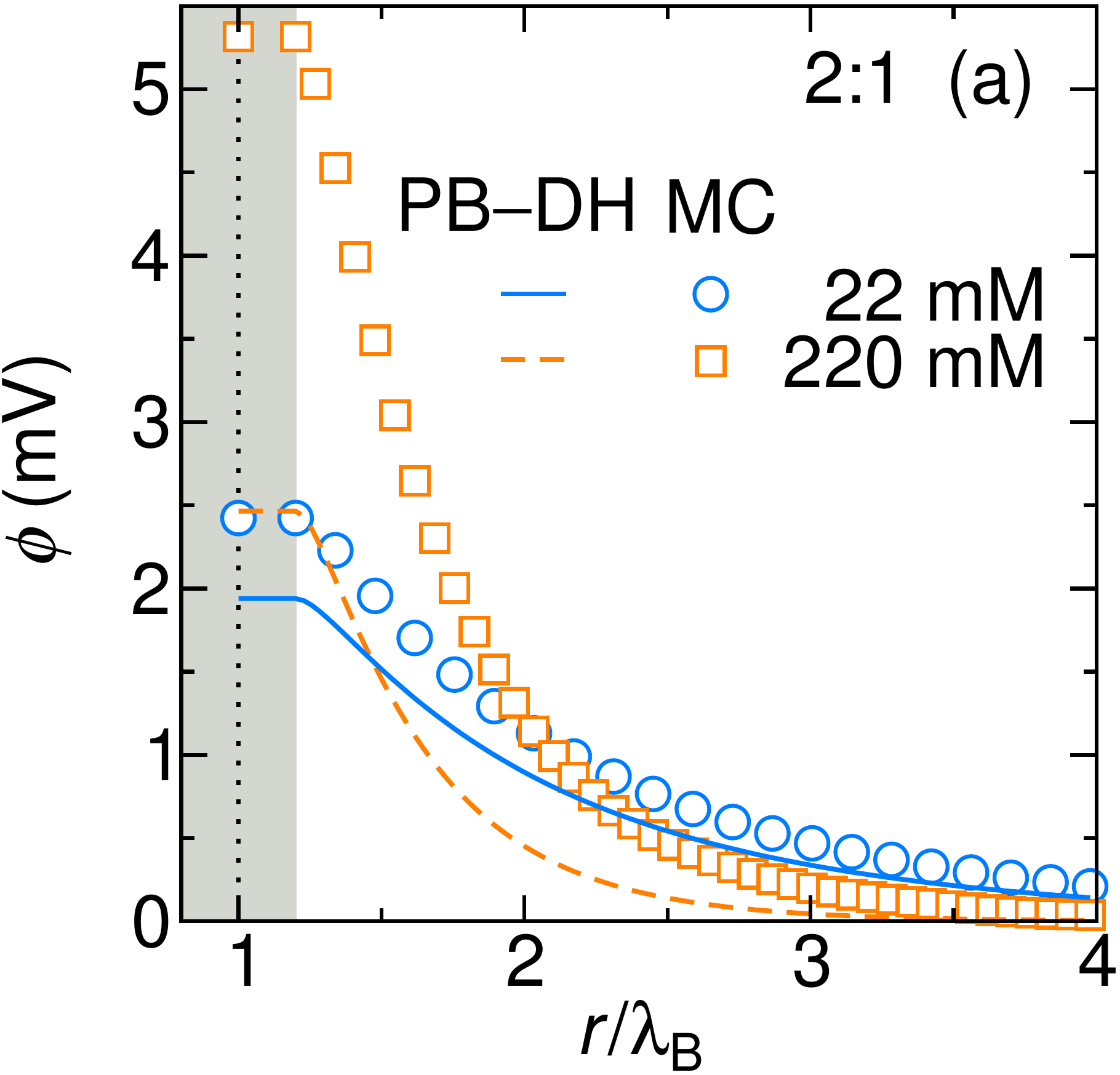} 
\end{center}\end{minipage}\hspace{1ex}
\begin{minipage}[b]{0.214\textwidth}\begin{center}
\includegraphics[width=\textwidth]{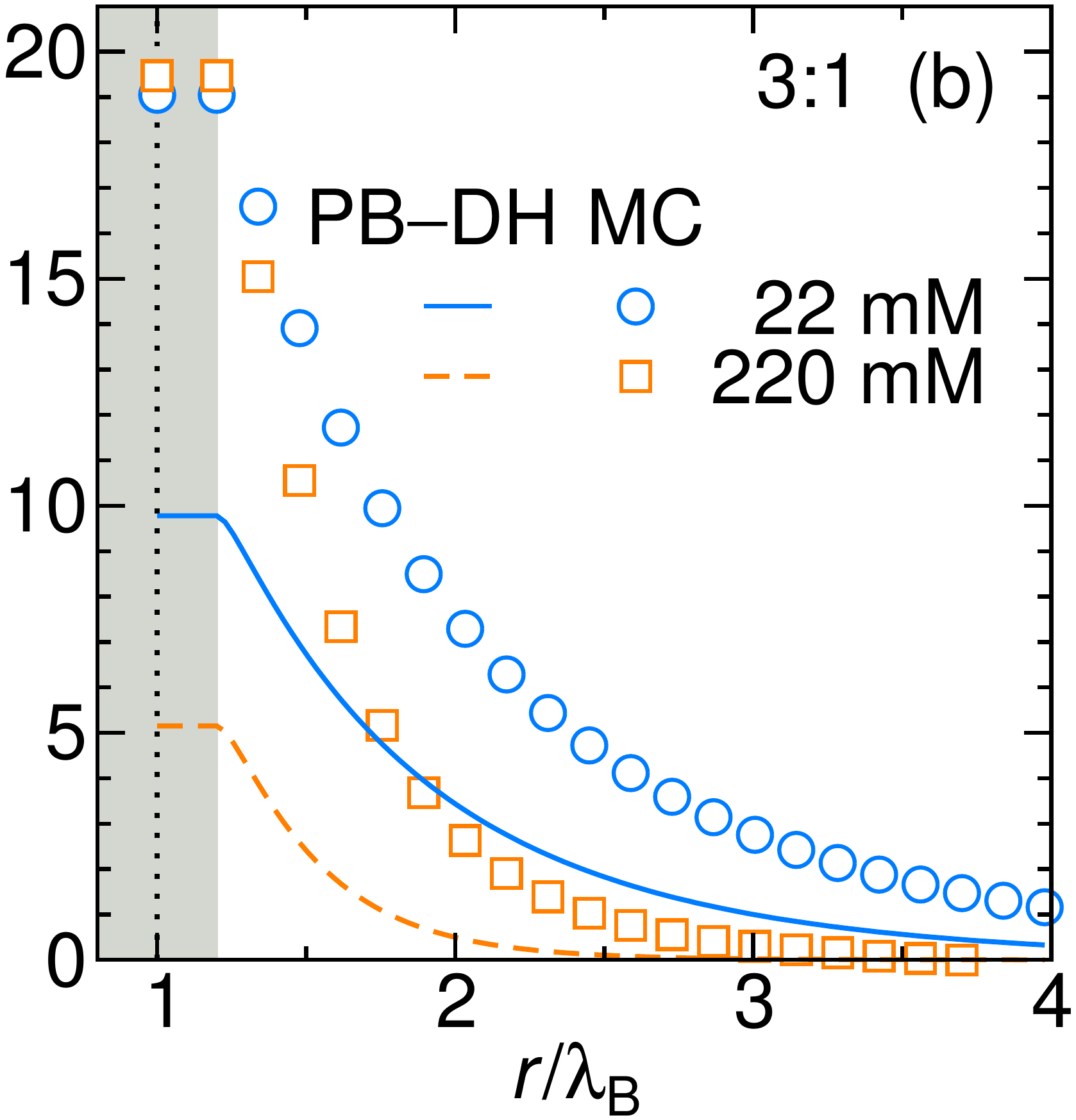} 
\end{center}\end{minipage}\vspace{1ex}
\begin{minipage}[b]{0.227\textwidth}\begin{center}
\includegraphics[width=\textwidth]{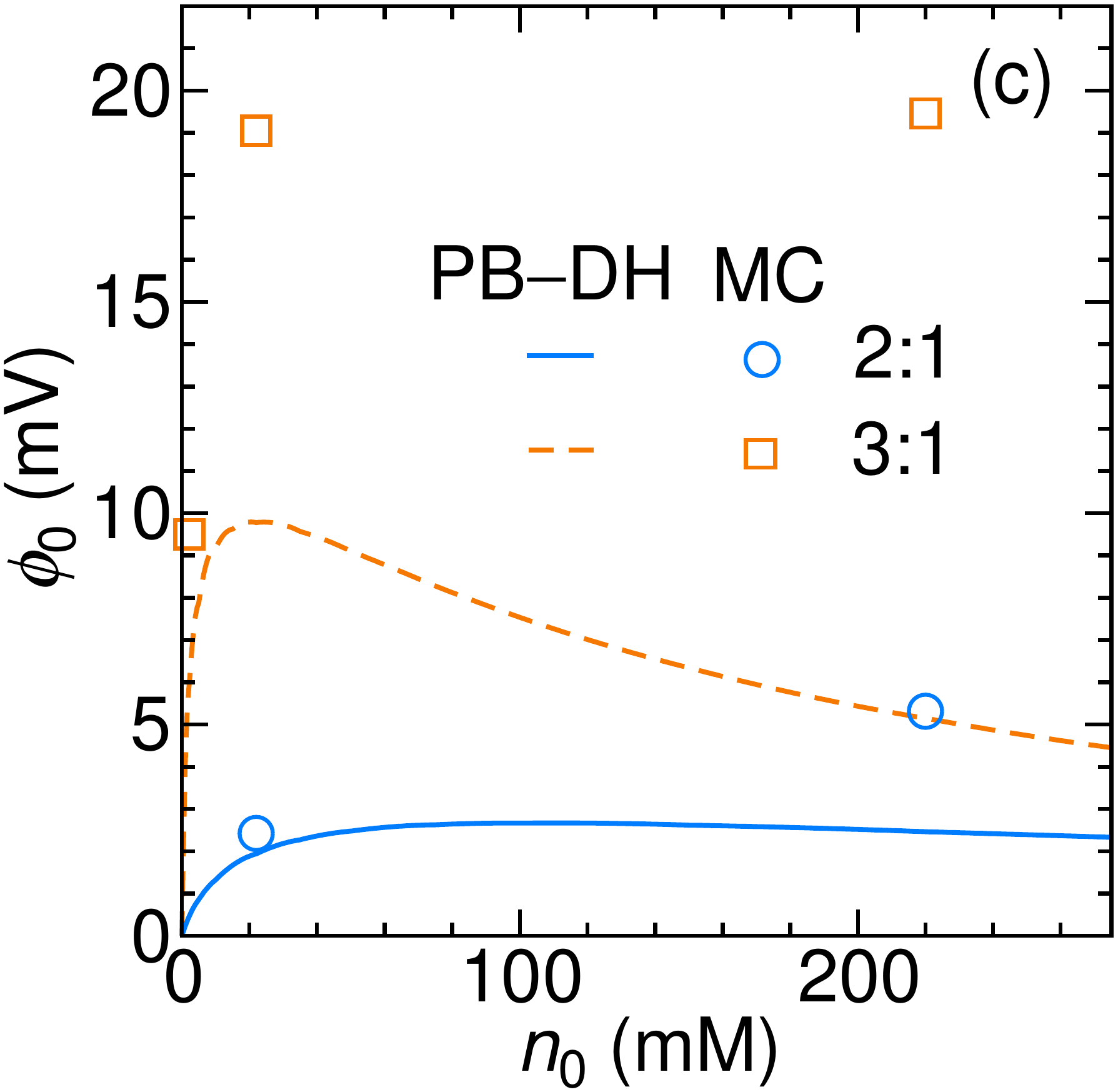} 
\end{center}\end{minipage}\hspace{1ex}
\begin{minipage}[b]{0.22\textwidth}\begin{center}
\includegraphics[width=\textwidth]{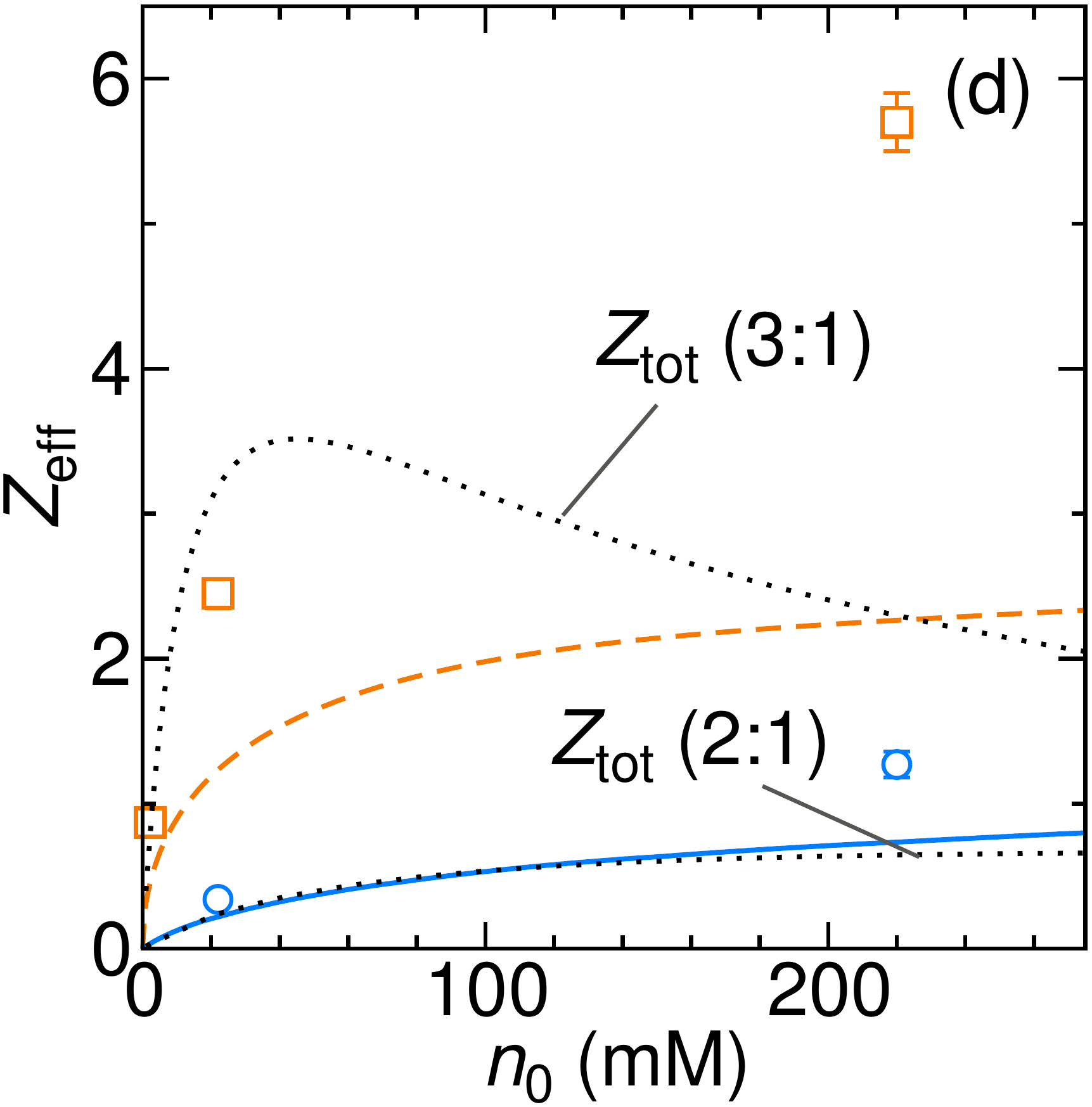} 
\end{center}\end{minipage}
\caption{Electrostatic potentials for 22 and 220~mM of (a)~2:1 and (b)~3:1 electrolytes.
(c)~Surface electrostatic potential as a function of concentrations of 2:1 and 3:1 electrolytes as obtained from the PB--DH theory (lines) and MC simulations (symbols). 
(d)~Effective charge $Z_\trm{eff}$ of the nanoparticle evaluated from fitting \Eq~(\ref{eq:DH-Zeff}) to the potential curves [same legend as in~(c)]. In addition, the total accumulative charge $Z_\trm{tot}$ from the B--DH theory is plotted by dotted curves.
}
\label{fig:pot}
\end{center}\end{figure}

In \Fig~\ref{fig:pot}a and b we plot the electrostatic potentials generated by asymmetric electrolytes. While PB--DH gives satisfactory agreement at 22~mM of 2:1 salt, it 
becomes poorer at 220~mM, where the deviation reaches a factor of 2. The situation significantly worsens for 3:1 case.
The surface potential $\phi_0$ as a function of concentration is plotted in panel~(c). 
The theoretical prediction, which is now only qualitative, predicts non-monotonic behavior. The surface potential first rapidly rises with concentration due to increased adsorption of ions. At larger concentrations, the rise of the adsorption slows down with increasing concentration due to electrostatic repulsion of already adsorbed ions. Additionally, increasing the salt concentration increases also the screening of the electrolyte, which eventually leads to a drop in the surface potential at high concentrations.
Whereas PB--DH yields satisfactory agreement for the 2:1 case (deviating by factor of 2 from MC at large concentrations), it fails considerably for the 3:1 case.
As predicted by the MC simulations, a 3:1 electrolyte creates approximately 20~mV of surface potential in the  range of 20--220~mM. This is comparable to the specific-adsorption model discussed in the previous section.

We now fit the DH theory [\Eq~(\ref{eq:DH-Zeff})] to the long-distance potential, which gives us the effective charge $Z_\trm{eff}$, shown in~(d).
Contrary to the specific-adsorption model in the previous section, the effective charge in this case is notably a non-linear function of concentration. It first shows a rapid increase with the concentration that turns into a more gradual trend at higher concentrations. Consistently with the results for $\phi_0$ in (c), the PB--DH theory underestimates the effective charge. 
Similarly as in the previous section, the total cumulative charge $Z_\trm{tot}$ from the B--DH approach is very similar to $Z_\trm{eff}$ from MC and PB--DH, with an exception for high concentrations of the 3:1 electrolyte. 

The last plot is revealing an immense influence of the valency asymmetry on the effective charge. 
According to the MC result, a neutral nanoparticle of a radius 1$\lB$ gains an effective charge of around 1~$e_0$ at 220~mM of 2:1 electrolyte, and an impressive 6~$e_0$ in a 3:1 electrolyte of the same concentration.
Here we note that the expected effective charge scales with an increasing nanoparticle size faster than its surface, since, as we have seen in \Fig~\ref{fig:size}, larger particles adsorb ions more effectively due to their higher polarizability. In the limit of large nanoparticle sizes, we then expect $Z_\trm{eff}\sim a^2$. That means that in the case of a polydisperse solution of particle sizes, larger ones gain significantly larger charges than smaller ones.



The presented model points to a practical relevance in the physical chemistry, namely
the build-up of an electric double layer even in the absence of surface charge, solely because of the difference in cation and anion concentrations in the surface vicinity.
The so-called ``zero surface-charge double layer'', a concept introduced by theoretical models
a few decades ago,~\cite{dukhin1982problem, derjaguin1987role} helped to interpret several experimental facts, such as electrokinetic effects of uncharged colloids.~\cite{dukhin, bazant2010induced, bazant2004induced}
A charged nanoparticle surface enhances its chemical reactivity and consequently has a
strong impact on its growth.~\cite{sylvestre2004surface}
In reality, metal nanoparticles can also possess an intrinsic charge. Partially because nanoparticles can be contaminated with various compounds from electrolytes and oxidized material.~\cite{mucalo2001electric, sylvestre2004surface}
On the other hand, some syntheses techniques of gold nanoparticles (\eg, pulsed laser ablation) lead to partial oxidation (3.3--6.6\%~\cite{muto2007estimation}) of surface atoms, forming a pH-dependent equilibrium of Au--OH/AuO$^{-}$ terminal groups, which thus contribute to the overall negative charge of gold nanoparticles. 


\section{Conclusions}

In this study, we revisited a continuum electrostatics problem of image-charge interactions and 
applied it to a model of a metal nanoparticle, featuring a high dielectric interior and hence high polarizability. 
We compared the predictions of various theoretical approaches, differing in their mathematical complexity and applicability regimes, with Monte Carlo simulations.

Focusing first on the case of symmetric electrolyte, we found very good agreement of the theoretical approaches and MC simulations. Here, the polarizability effects lead to sizable ion accumulation near the nanoparticle surface, which further depends on the ionic strength as well as on the nanoparticle size.
In addition, we investigated \chgA{how an asymmetry in the adsorption affinities for cations and anions influences their distributions}. We separately considered two different kinds of asymmetries, in one case stemming from an additional specific adsorption potential to one ionic species, and in the other case stemming from an asymmetric electrolyte (\ie, 2:1 and 3:1).
The asymmetries, which give rise to asymmetric distributions of ionic profiles, engender a net electrostatic potential and an effective charge of the nanoparticle.
Here, even the most simple approaches that neglect the generated potentials can nevertheless very satisfactorily predict local ion densities (\ie, in the surface vicinity). Of course, at larger distances, where ions tend to neutralize the accumulated charge, it is necessary to invoke a Poisson--Boltzmann description with implemented image-charge corrections.
For very high charge asymmetries, such as in a 3:1 electrolyte, the theories face difficulties when compared with \chgA{the} ``exact'' solutions of MC simulations.
\chgA{The difficulties may be associated with correlation effects between multivalent ions, which are not captured within our theoretical framework.
Still, the theories} are able to capture the qualitative behavior considerably well and thus help to elucidate basic principles of electrostatics of metal nanoparticles in electrolyte solutions.

Finally, we need to be aware of various conceptual challenges that occur
in such systems containing metal-like particles in aqueous solutions.
Due to high ionic adsorption affinities, the surface details become very important.
This is in stark contrast to low-dielectric macromolecules, where ions are typically repelled from the surfaces, and therefore their molecular structure becomes less relevant.
One of such details is for instance the exact geometry of the nanoparticles, which typically possess a well-defined atomic arrangement (\eg, resembling the face-centered cubic structure,~\cite{petkov2005structure}) and seems to be critical for the nanoparticle's activity.~\cite{mahmoud2013enhancing}
A deeper understanding of fine details of metal nanoparticles calls for approaches beyond the idealized continuum model. In this context, in particular atomistic models that take the granularity of the nanoparticle surface and solvent into account are nowadays becoming the focus of  sophisticated simulation approaches.~\cite{heikkila2012atomistic, chen2014molecular, li2015molecular}

\section*{Appendix: Green's function near a metal sphere}
We decompose the Debye--H\"{u}ckel Green's function $u^\trm{DH}(\Av r, \Av r')$ into the direct and indirect part [similarly as the Coulomb Green's function given by \Eq~(\ref{eq:u_image})],
\begin{equation}
u^\trm{DH}(\Av r, \Av r')=u_0^\trm{DH}(\Av r, \Av r')+u_\trm {im}^\trm{DH}(\Av r, \Av r')
\label{eq:uDH_image}
\end{equation}
We now express it in spherical coordinates with the center of the sphere located in the origin of the coordinate system, that is, $\Av r(r, \theta,\varphi)$ and $\Av r'(r', \theta',\varphi')$. A multipole expansion of the direct part yields a form,~\cite{boschitsch1999fast}
\begin{equation}
u_0^\trm{DH}(\Av r, \Av r')=\frac{8\kappa}{4\pi\varepsilon\varepsilon_0}\sum_{lm} i_l(\kappa r_<)k_l(\kappa r_>)Y_{lm}(\theta',\varphi')Y_{lm}^*(\theta,\varphi)
\label{eq:u0s}
\end{equation}
Here, $r_<$ and $r_>$ correspond respectively to the smaller and the larger radial value among $r$ and $r'$.
The functions $Y_{lm}$ are spherical harmonics, the asterisk denotes the complex conjugate value, and the spherical modified Bessel functions are defined by \Eq~(\ref{eq:bessel}).
The summation in \Eq~(\ref{eq:u0s}) runs over integer values $l=0,1,2\ldots$ and $m=-l,\ldots,+ l$.

A general {\it Ansatz} for the second term in \Eq~(\ref{eq:uDH_image}) is~\cite{spherical2, javidpour2018pre}
\begin{equation}
u_\trm{im}^\trm{DH}(\Av r, \Av r')=\sum_{lm}A_{lm}k_l(\kappa r)Y_{lm}^*(\theta,\varphi).
\end{equation}
In order to determine coefficients $A_{lm}$, we apply two boundary conditions.
The first condition requires that the potential on the surface of the metal sphere with a radius $a$ is constant, that is, independent of the solid angle $\Omega$
\begin{equation}
\frac{\partial u^\trm{DH}(\Av r,\Av r')}{\partial \Omega}|_{r=a}=0
\label{eq:BC1}
\end{equation}
which leads to
\begin{equation}
A_{lm}=-\frac{8\kappa }{4\pi\varepsilon\varepsilon_0}\frac{i_l(\kappa a)}{k_l(\kappa a)}k_l(\kappa r') Y_{lm}(\theta',\varphi')\quad\trm{for } l,m\ne0.
\end{equation}
Note that for $l=m=0$, $Y_{00}=1/\sqrt{4\pi}$, and consequently $\partial Y_{00}/\partial\Omega=0$, thus trivially satisfying \Eq~(\ref{eq:BC1}). For that reason, $A_{00}$ has to be determined by an additional boundary condition. Since the sphere is electrically isolated, it is overall charge neutral. We apply the Gaussian law, $\oint \Av E\cdot \rmd \Av S = 0$, where we integrate the electric field $\Av E=-e_0 \nabla u^\trm{DH}$ over the sphere surface. This provides the second boundary condition
\begin{equation}
\int\frac{\partial u^\trm{DH}}{\partial r}\Big|_{r=a}\sin\theta\rmd\theta\rmd\varphi=0
\label{eq:BC2}
\end{equation}
where we integrate over the entire solid angle.
Using the identity $\int Y_{lm}(\theta,\varphi)\sin\theta\rmd \theta\rmd\varphi=\sqrt{4\pi}\,\delta_{l0}\delta_{m0}$, which eliminates all the terms but $l=m=0$ when performing the integration in \Eq~(\ref{eq:BC2}), provides the remaining coefficient
\begin{equation}
A_{00}=-\frac{8\kappa }{4\pi\varepsilon\varepsilon_0}\frac{i'_0(\kappa a)}{k'_0(\kappa a)}k_0(\kappa r') Y_{00}(\theta',\varphi')
\end{equation}
The self-energy of a monovalent ion then follows as $w_0^\trm{DH}=(1/2)e_0 u^\trm{DH}(\Av r,\Av r)$, which is \Eq~(\ref{eq:DHw0}) in the main text.

\section*{Conflicts of interest}
There are no conflicts of interest to declare.

\section*{Acknowledgments}
The authors thank Won Kyu Kim, Victor G.\ Ruiz L\'{o}pez, Matthias Ballauff, and Rudolf Podgornik for useful discussions.
This project has received funding from the European Research Council~(ERC) under the European Union's Horizon 2020 research and innovation programme (grant agreement n$^\circ$~646659-NANOREACTOR).
The simulations were performed with resources provided by the North-German Supercomputing Alliance~(HLRN).

\bibliography{bibliography-sphere}
\bibliographystyle{rsc}

\end{document}